# Stealth Communication: Zero-Power Classical Communication, Zero-Quantum Quantum Communication and Environmental-Noise Communication

Laszlo B. Kish[a)]

Texas A&M University, Department of Electrical Engineering, College Station, TX 77843-3128, USA;   email: Laszlo.Kish@ee.tamu.edu

**Abstract**. An alternative physical way of communication, *communication by the inherent background noise*, is proposed which does not need net energy transfer in the information channel. The communicator devices do dissipate energy; however, they do not emit net energy into the channel, instead of that, they modulate the parameters of inherent spontaneous fluctuations in the channel. The method can use two different mechanisms, thermal noise (Johnson-Nyquist noise) for classical communication, and vacuum fluctuations/zero point energy (quantum uncertainty noise) for quantum communication. The strongest advantage of the method that this is apparently the *most hidden* (*stealth*) way of communication, because it is using the inherent background noise for communication, therefore it is extremely difficult or impossible to discover its presence. With proper wave-based arrangements and specific conditions, the sender and the receiver can easily detect eavesdropper activities, so that the eavesdropper is detected as soon as she extracts a single bit of information, thus the security of the method is comparable to the security of quantum communication/quantum key distribution schemes. Finally, concerning practical applications, environmental noise, out of the fundamental/inherent fluctuations, can also be used for this kind of communication provided that is sufficiently stationary.

**Keywords:** Stealth communication; eavesdropper detection; thermal noise; zero-point energy; vacuum fluctuations; communication by noise.

---

In this Letter we show that it is possible to communicate via an information channel without putting *net energy* into the channel. This surprising result is achieved by the modulation of the parameters of *equilibrium-thermal* or *zero-point-quantum* fluctuations in the channel. Thus this communication is using the inherent energy fluctuations, which is called channel noise in the literature, to carry the information.

First of all, an important fact is emphasized, namely that *this method is very different from the so-called energy-free communication.* There have been some discussion and debate by prominent scientists about the possibility of energy-free communication [1,2]. However it is important to note that the *methods and issues presented in this paper are very different* because:

- The method presented in this paper is utilizing and detecting the *equilibrium-thermal* and/or *zero-point-quantum* fluctuations in the information channel. All the debated arrangements mentioned above make strong efforts to *suppress* these fluctuations *below the detection limit* where they do not disturb the information processing any more.

- Even if there is *no net energy* communicated in the channel, there is a *huge energy*, much more than *kT/bit* dissipated in the communicator devices at the sender and receiver's side; unlike in the debated arrangements mentioned above, where there is an effort to regain the energy spent. We do *not* make any effort to regain any type of energy used to run the system.

In conclusion, the debate [1,2] about energy-free communication, though intriguing, is irrelevant here.

After realizing the fact that the inherent fluctuations can be used for communication, the realization and the various solutions are self-evident and straightforward. From the many possibilities, we show only a few examples in this paper. Without the restriction of generality, we demonstrate the idea by voltage fluctuations described by the *fluctuation-dissipation theorem* (FDT), in the classical physical and the quantum physical limits, respectively. According to the FDT [3-8], the generalized Johnson-Nyquist formula of equilibrium thermal noise in the voltage can be written as:

$$S_u(f) = 4[N(f,T) + 0.5]hf \operatorname{Re}[Z(f)] \tag{1}$$

where $S_u(f)$ is the power density spectrum of the voltage noise on the open-ended impedance $Z(f)$; $\operatorname{Re}[Z(f)]$ is the real part of the impedance; and $h$ is the Planck constant. The Planck number $N(f,T)$ is the mean number of $hf$ energy quanta in a linear harmonic oscillator with resonance frequency $f$, at temperature $T$:

$$N(f,T) = [\exp(hf/kT) - 1]^{-1} \quad , \tag{2}$$

which is $N(f,T) = kT/(hf)$ for the classical physical range $kT \gg hf$ so that the mean energy in the oscillator is $kT + 0.5hf$, where the second term is the zero-point energy. The familiar thermal noise voltage formula is valid in this classical limit, that is for the low frequency range $f \ll kT/h$, where the first energy term dominates:

$$S_{u,class}(f) = 4kT \operatorname{Re}[Z(f)] \tag{3}$$

In the quantum limit, which is the high frequency limit $f \gg kT/h$, the zero-point fluctuations can be measured with a linear voltage amplifier:

$$S_{u,quan}(f) = 2hf \, \text{Re}[Z(f)] \quad . \tag{4}$$

Although the proper physical interpretation of Eq. 4 is somewhat debated [5,7,8] (whether the observed noise is an explicit zero-point fluctuation or if it is the manifestation of the uncertainty principle, etc.), the validity of Eq. 4 for linear voltage amplifiers in the quantum limit is commonly accepted, confirmed by experiments and applied in engineering [4,6,7]. Therefore, because the practical existence of this quantum noise phenomenon *in the voltage* is a fact, the debate mentioned above is irrelevant for the present paper, and here we can take Eq. 4 as the voltage noise of impedances in the quantum range. It is however important to mention [6] that this noise is zero in a photocell or other detectors which are detecting particle number instead of field amplitude and measure only the Planck radiation. Thus, we always envision a *linear voltage amplifier* input at the receiver's side when we discuss *quantum noise* examples in the present paper.

The *general outline* of the zero-signal-power communication system *for classical information* is shown in Fig 1. The $f \ll kT/h$ classical physical condition holds and the impedance generating the thermal noise of the channel (see Eq. 3) is modulated by the sender. The receiver detects the signal, which is the modulation of the statistical properties such as spectrum, hidden correlations, etc. in the thermal noise. Though controlling the communication devices requires energy, the signal energy pumped into the channel and the net energy transfer between the sender and receiver is zero. Note, that with alternative, wave-based solutions, the sender can modulate the reflection coefficient of the channel at his end and introduce time-correlations in the noise (see below).

An example of a simple realization of the *zero-signal-power classical communicator* can be seen in Fig. 2. The information-bit is represented by the choice of two different impedance-bandwidths and the two corresponding thermal noise spectra (see Equation 3) provided by the two different capacitors at the sender's side of the channel, which is made of two wires of short length $L << c_p / f$, where $c_p$ is the propagation velocity of electromagnetic fields in the cable. The noise analyzer can measure power density spectra, zero-crossing frequency autocorrelation or other related quantity. The advantage of the method shown in Fig. 2 is that for everybody, except for an educated eavesdropper, it looks like there is no communication in the channel because only the usual background noise (thermal noise) is present and the signal power is zero. However the educated eavesdropper with proper equipment and settings can observe that the high-cut-off frequency of the thermal noise is switching between two distinct values.

The *general outline* of the zero-signal-power communication scheme *for quantum information* is shown also in Figure 3, with the condition $f >> kT / h$. The impedance determining the spectral properties of vacuum fluctuations in the channel (see Eq. 4) is modulated by the sender. Though controlling the communication devices require many energy quanta, no quanta are sent through the information channel and the energy transfer between the sender and receiver is zero. Similar solution can also be achieved by two coupled quantum systems one at the sender's side and one at the receiver's side, respectively. Note that concerning *electronic* solutions there is an apparent disadvantage compared to the classical schemes. The $kT << hf_c$ condition requires amplifying voltages at very high frequencies or being in a very low temperature environment. Therefore this kind of *electronic quantum communication* seems to be more relevant for communication in space where the background thermal noise temperature is around 2.7K. However heterodyne techniques such as TerraHertz or optical phase sensitive techniques may still be relevant.

The example outlined in Fig. 3 can also be used as a simple realization of the *zero-quantum quantum communicator* provided that the $kT << hf_c$ condition holds, $f_c$ is the upper cut-off frequency. The information-bit is represented by the choice of two impedance-bandwidths and the two corresponding uncertainty-noise spectra, provided by the two different capacitors at the sender's side of the wire-channel. The advantages and disadvantages of this system are the same as those of the classical system described above. An extra problem is the need of using low temperature and/or high frequency.

At considering *wave-based realizations*, for the sake of simplicity and practical relevance, we restrict ourselves to the case of classical thermal noise, though considerations about the quantum case would be similar. Wave-based arrangements need longer cables than half of the wavelength corresponding to the bandwidth, $L >> c_p / f$. They can also be realized with the bandwidth modulation systems shown in Fig. 2, however modulation of the reflection can be a more advanced option. Figure 4 shows a wave-based arrangement with a coaxial cable and matching resistor at the receiver's end the modulation of the reflection at the sender's end. The detection of the information at the receiver's side needs a correlator (a multiplier and an averaging circuit) and a delay line to determine the value of the autocorrelation function at time equal to the time of return flight of the wave along the cable. The three stages of the switch at the sender's side represent a three-level information system: 1 no reflection (no correlation); 2 positive reflection (positive correlation); 3 negative reflection (anti correlation). Observing the reflection properties via the autocorrelation function provides extraordinary sensitivity to detect possible eavesdropper activities because any activity along the line will cause reflections at different time delays.

The presence of an eavesdropper can be detected in a wave-based arrangement with two parabolic antennas. The arrangements at the sender's and receiver's sides are the same as in Fig. 4, however this time the signal propagates in the space. When used as a classical communicator, in *thermal equilibrium*, this arrangement has the highest sensitivity to detect an eavesdropper. The eavesdropper, by using a tilted reflector, can couple out a part of the *signal-carrying-noise* from the communication channel between the two antennas, however she will always induce excess noise, when she uses a matched resistor closing. Otherwise, if she leaves the antenna electrodes ends open or shorted, she is causing excess reflection maxima/minima in the correlation function at the receiver's device. Because the out-coupled *signal-carrying-noise* has the same wave intensity as the *useless thermal noise* of the external space, out-coupling has to be significantly strong to have a reasonably good statistics of the detection. What is most important, the receiver will observe at least as large change on the autocorrelation function as the signal detected by the eavesdropper. Thus, as soon as the eavesdropper detects a bit with a certain probability, the receiver will detect the eavesdropper's presence with the same probability. Therefore while the eavesdropper detects a single bit, the receiver detects the presence of the eavesdropper. Even though it is a classical communicator, this property *of the thermal equilibrium situation* is superior to known quantum key distribution schemes where the eavesdropper's presence can be detected only after she extracts a great number of bits by using a quantum amplifier.

However, the above described superior characteristics of the classical communicator hold *only in thermal equilibrium*. It has been shown by Henry Taylor [9] that an eavesdropper using sufficiently cooled nonreciprocal devices (Faraday isolator) would be able to extract information while staying hidden. Thus, in the *classical communicator* case, the absolute security holds only in situations where the eavesdropper is unable to use devices with lower temperature than the temperature of the communicator system. Finally, if this communicator is operated in the *quantum limit*, the temperature has no role in determining the working conditions because the communication takes place via parametric modulation of the vacuum fluctuations of the ground state. Therefore, the study of the visibility of the eavesdropper needs a quantum treatment of the situation. Taking into the account the properties of the quantum noise of resistors due to zero-point fluctuations [3-5,8] (and the fact that there is no lower noise energy than the zero-point energy), we suspect that the visibility of the eavesdropper will be similar to that of the classical communicator in the case of thermal equilibrium, that is the eavesdropper can probably be discovered after extracting a single bit of information.

Finally we note that, if *stationary environmental noises* dominate the channel and the sender can modulate the bandwidth, reflection or some statistical property of noise, the zero-signal-power communication can be executed in the same ways as described above. The properties will be similar but the noise will be stronger than the fundamental limits.

In this Letter, we outlined a new way of communication with zero signal power. The main advantage of this method, as compared to other kinds of classical and quantum communication (quantum key distribution), is that it is stealth because only a background noise can be detected in the channel. At specific conditions, in thermal equilibrium, secure communication can be made even with classical information.

**Acknowledgement**

This paper is the subject of a TAMU patent disclosure submitted on August 18, 2005. Henry Taylor (TAMU) has shown how to break the encryption in the microwave based arrangement with cooled devices, out of thermal equilibrium. The author is grateful also to the following colleagues for valuable discussions: Peter Hrasko, Janos Bergou, Marlan Scully, Attila Geresdi, Julio Gea-Banacloche, Bob Biard, Suhail Zubairy.

**Figure caption**

**Figure 1.**

Zero-signal-power classical communication. General outline of the zero-signal-power communication scheme for classical information.

**Figure 2.**

Simple realization of the zero-signal-power classical or quantum communicator. The information-bit is represented by the choice of two different impedance-bandwidths and the two corresponding thermal noise spectra. The noise analyzer can measure power density spectra, zero-crossing frequency autocorrelation or other related quantity.

**Figure 3.**

Zero-quantum quantum communication. General outline of the zero-signal-power communication scheme in the quantum limit.

**Figure 4.**

Wave-based arrangements with a coaxial cable and matching resistor at the receiver's end the modulation of the reflection at the sender's end.

**Figure 5.**

Wave-based arrangements with microwaves, two parabolic antennas and the reflection modulation technique shown in Fig 5. While the eavesdropper detects a single bit, the receiver detects the presence of the eavesdropper.

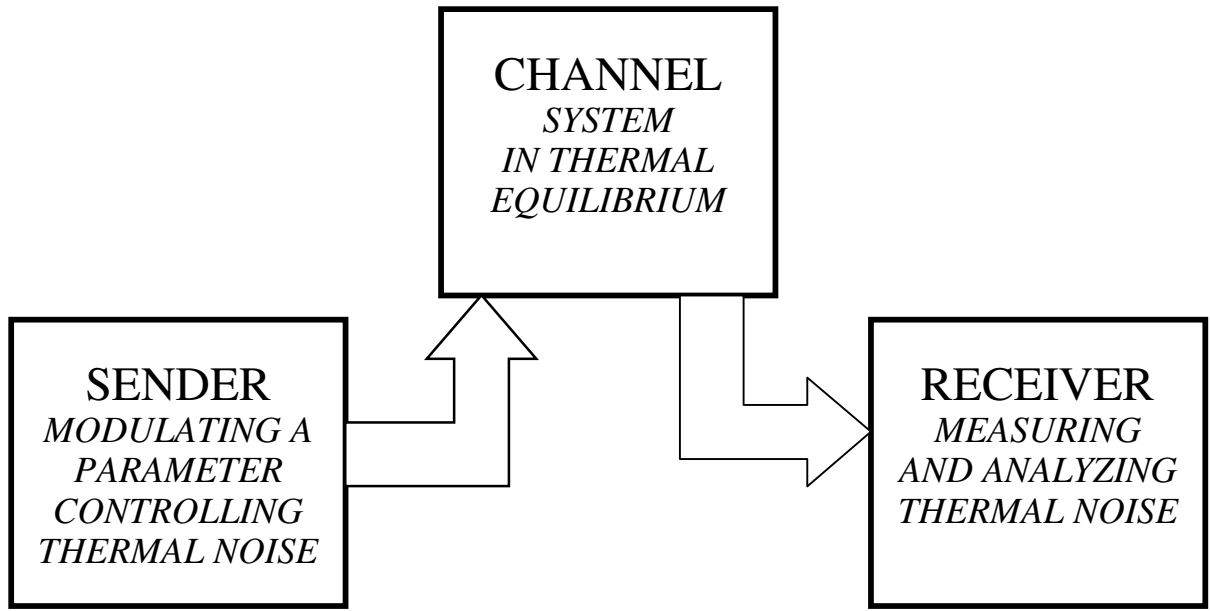

**Figure 1.**

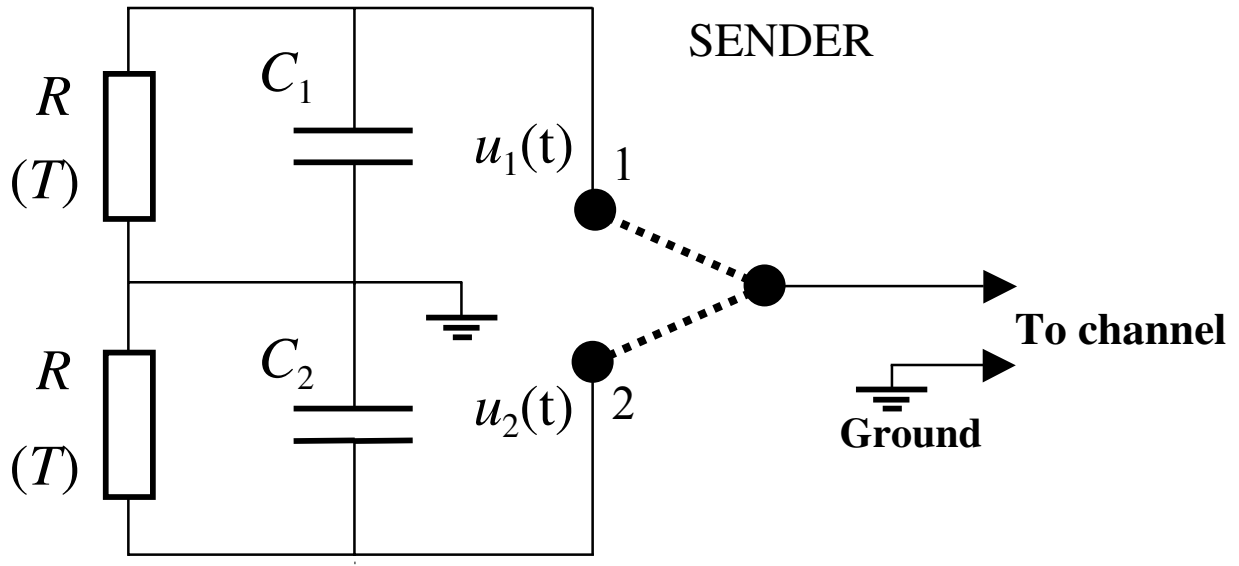
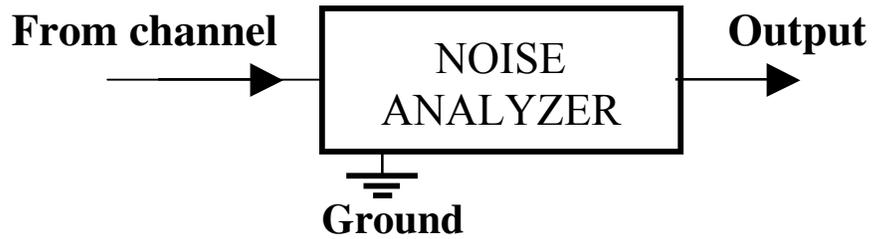

**Figure 2.**

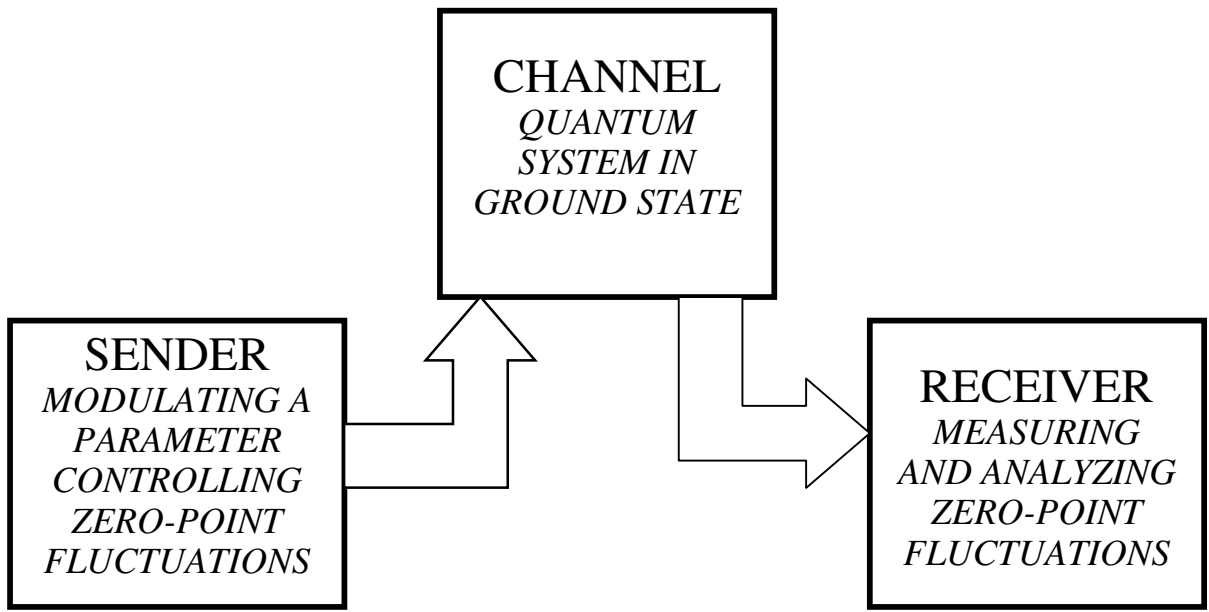

**Figure 3.**

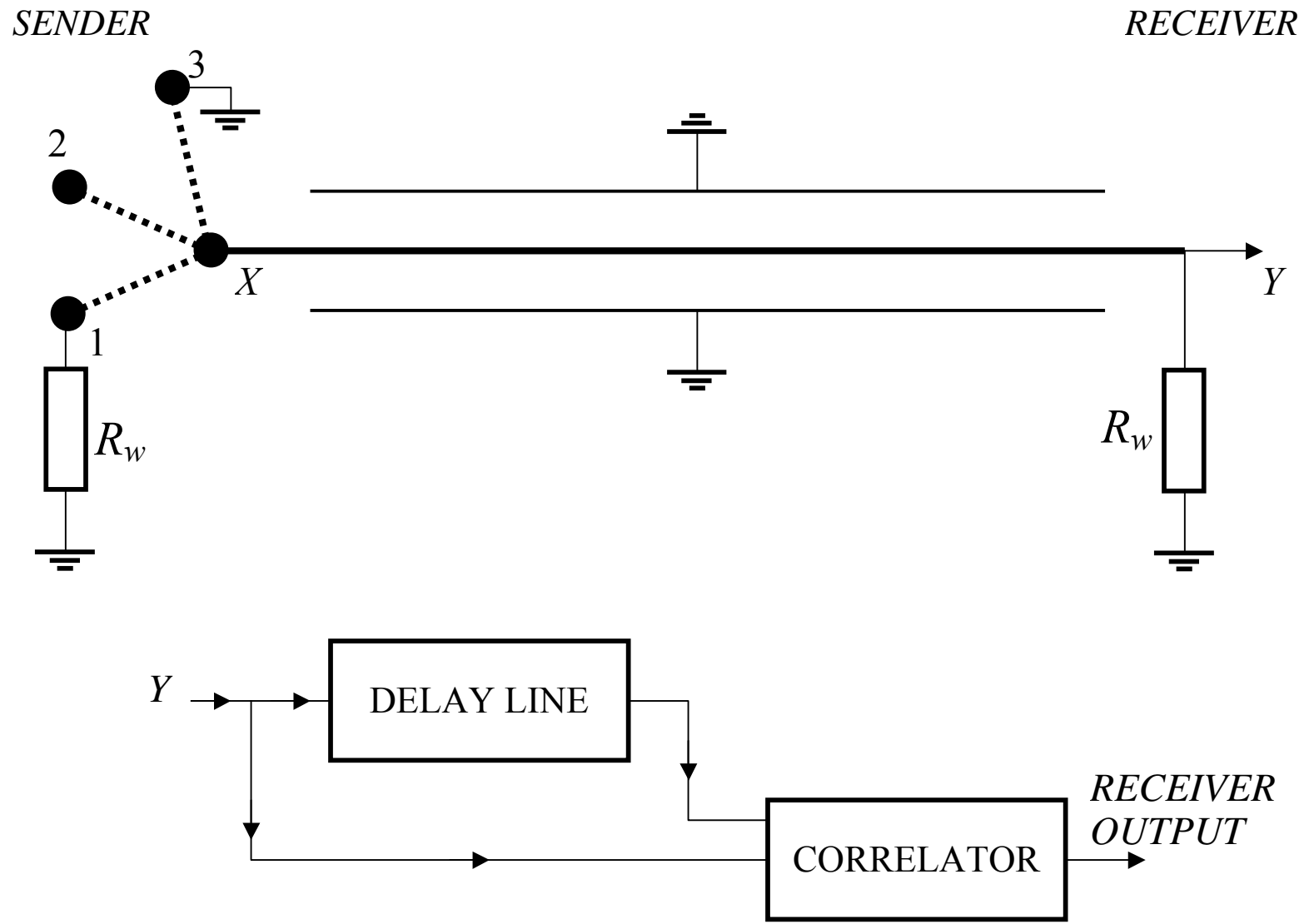

**Figure 4.**

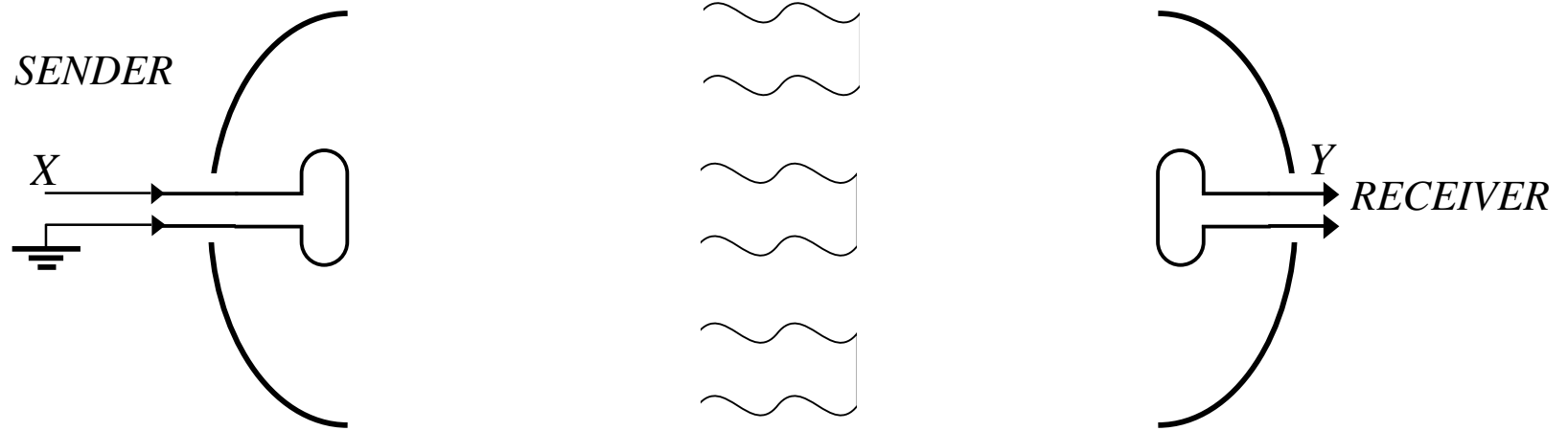

**Figure 5.**